\documentclass[onecolumn,preprintnumbers,amsmath,amssymb,showpacs,aps,preprint,prb]{revtex4}

\usepackage{graphicx}	
\usepackage{dcolumn}	
\usepackage{bm}		
\usepackage{color}
\usepackage{colordvi}
\usepackage{epsfig}

\newcommand{\rr}{{\bf r}}

\begin{document}

\title{Structural precursor to freezing: An integral equation study}

\author{Joseph M. Brader}
\affiliation{Fachbereich Physik, Universit\"at Konstanz, D-78457 Konstanz, 
Germany}

\date{\today}

\begin{abstract}
Recent simulation studies have drawn attention to the shoulder which forms in the 
second peak of the radial distribution function of hard-spheres at densities close 
to freezing and which is associated with local crystalline ordering in the dense 
fluid. We address this structural precursor to freezing using an inhomogeneous 
integral equation theory capable of describing local packing constraints to a high 
level of accuracy. 
The addition of a short-range attractive interaction leads to a well known broadening 
of the fluid-solid coexistence region as a function of attraction strength. 
The appearence of a shoulder in our calculated radial distribution functions is 
found to be consistent with the broadened coexistence region for a simple model 
potential, thus demonstrating that the shoulder is not exclusively a high density 
packing effect.    
\end{abstract}


\maketitle

\section{Introduction}
The hard-sphere system has been of central interest in the theory of fluids since the 
early work of Boltzmann and is still an important model system in statistical physics. 
At low and intermediate densities the equilibrium pair structure and thermodynamics are
theoretically well understood and the model can be regarded as solved, in the sense that
the numerical results agree with simulation to an accuracy satisfactory for most
applications. 
At high density the system undergoes a first-order liquid-solid transition to an FCC 
crystal \cite{alder,hoover} the theoretical description of which remains an open 
problem.
The challenge for theoretical treatments of the freezing transition is to obtain a unified
description of both the amorphous fluid and the symmetry broken solid. The
density functional theory of freezing provides a pragmatic approach to 
this problem in which the solid is regarded as a highly inhomogeneous fluid 
\cite{bob_review}.
By employing a seperate treatment of the fluid and solid branches of the free energy 
(crystal symmetry is input to the theory by hand) modern density functional approximations 
are able to locate the coexisting densities of the hard-sphere freezing transition in close 
agreement with simulation \cite{tarazona,white_bear}. 
Nevertheless, a unified theory in which the crystal symmetry 
emerges spontaneously upon increasing the density is still lacking.  
\par
One possible method of gaining insight into the emergence of crystalline order is 
to investigate in detail the microscopic structure of the fluid state at densities close 
to, but below, the freezing transition. 
Recent simulations performed within this density range 
suggest the existence of precursor structures to freezing in the high density liquid 
state of hard-spheres in two and three-dimensions. 
It has been demonstrated \cite{truskett} using molecular dynamics simulations 
that the radial distribution function (RDF) of both the hard-disk and hard-sphere systems 
develops a shoulder feature in the second peak which first appears at packing fractions 
within roughly five percent of their respective freezing transitions. 
To facilitate visualization of configurations the 
analysis focused primarily on the hard-disk system
for which a signature four-particle hexagonal-close-packed configuration of disks was 
identified as being responsible for the observed shoulder in the RDF. 
This distinct structural motif reflects the next-nearest-neighbour ordering of the 
hexagonal domains which develop as the density is increased towards freezing. 
An analogous scenario was also proposed for the hard-sphere system 
but not analyzed in detail. 
A careful simulation study of the three-dimensional case \cite{snook} revealed six-membered 
ring configurations which flatten and increase in frequency close to freezing were 
identified as the relevant 
structural motif responsible for the shoulder observed in the simulation RDF. 
The existence of the shoulder has long been recognized and exploited by simulators as 
an empirical indicator for crystallization in hard particle systems, enabling expensive 
free-energy calculations to be avoided. 
As as result of the recent simulation studies \cite{truskett,snook} the microscopic origin of this
feature is now understood. 
Considering more general interparticle interactions, it is natural to inquire whether 
similar precursor structures also occur in systems with an attractive component to 
the pair potential. 
In particular, hard-spheres with short-range attraction 
present a liquid-solid coexistence region which broadens dramatically with increasing 
attraction strength, leading to coexistence between low-density fluid 
and high-density solid phases.  
The possible existence of local crystalline ordering at statepoints 
in the vicinity of this more general freezing boundary remains to be investigated and 
raises the question whether the RDF shoulder may also be present at low density 
statepoints far removed from the hard-sphere transition point.     
\par  
The most powerful theoretical technique for calculating the pair correlations and 
thermodynamics of classical fluids in equilibrium is the method of integral equations, 
formed by approximate closures of the Ornstein-Zernike equation \cite{hansen,caccamo}. 
For many model interaction potentials integral
equations provide highly accurate structural information at low and intermediate
coupling strength (although difficulties do remain in the vicinity of critical points,
when they occur \cite{brader_ijtp}).  
It is therefore surprising that none of the regularly applied integral equation 
theories for the pair structure are capable of describing the RDF shoulder
of hard-spheres and miss completely this well established feature 
of the equilibrium fluid pair structure, despite their admirable success at fluid 
statepoints removed from freezing. 
This failure raises the fundamental question whether 
the method of integral equations is capable, either in practice or in principle, of 
indicating the existence of a fluid-solid transition. 
Finding an integral equation theory which can correctly predict from first principles 
the detailed structure of the RDF in the high density liquid may 
thus shed some light on this important theoretical question \cite{bridge_footnote}.  
\par
In this paper we address the issue of precursor structures using inhomogeneous integral
equation theory. This class of theory is based upon the inhomogeneous Ornstein-Zernike
equation \cite{bob_review,hansen} in combination with closure relations between the 
inhomogeneous direct and
total correlation functions. In contrast to their homogeneous counterparts, which involve 
relations between pair functions, inhomogeneous integral equation theories work at 
the level of the triplet correlations. 
As the problem of crystallization and the appearance 
of precursor structures is associated with an increased local orientational 
ordering in the system, it is reasonable to suppose that increasing the orientational 
resolution of the theoretical treatment, i.e. working at the triplet rather than the 
pair level, will better capture the relevant physics. 
While a general theory of the emergence of crystalline order is still highly desirable 
(and we make no claims to provide this) we demonstrate that inhomogeneous integral 
equation theory predicts the occurance of an RDF shoulder in the vicinity of freezing 
and thus yields indirect evidence for the existence of underlying precursor structures.
By considering purely equilibrium fluid statepoints below the freezing transition 
we avoid the potential complications associated with metastable states.
\par
The remainder of the paper will be structured as follows: In Section \ref{theoretical} 
we  outline the phenomenology of the shoulder in the hard sphere RDF as found 
in simulation, briefly review existing integral equation approaches and  
introduce the inhomogeneous integral equation theory to be employed in this work.
In Section \ref{hardspheres} we present numerical solutions of the inhomogeneous theory 
for hard spheres and compare these with simulation data. 
In Section \ref{attraction} we consider the influence on the RDF of adding a 
short-range attraction to the hard-sphere pair potential. 
Finally, in Section \ref{discussion} we will discuss our results and give an outlook 
for future work.

\section{Theoretical approaches}
\begin{figure}
\newpage
\hspace*{-0.5cm}
\includegraphics[width=14cm,angle=0]{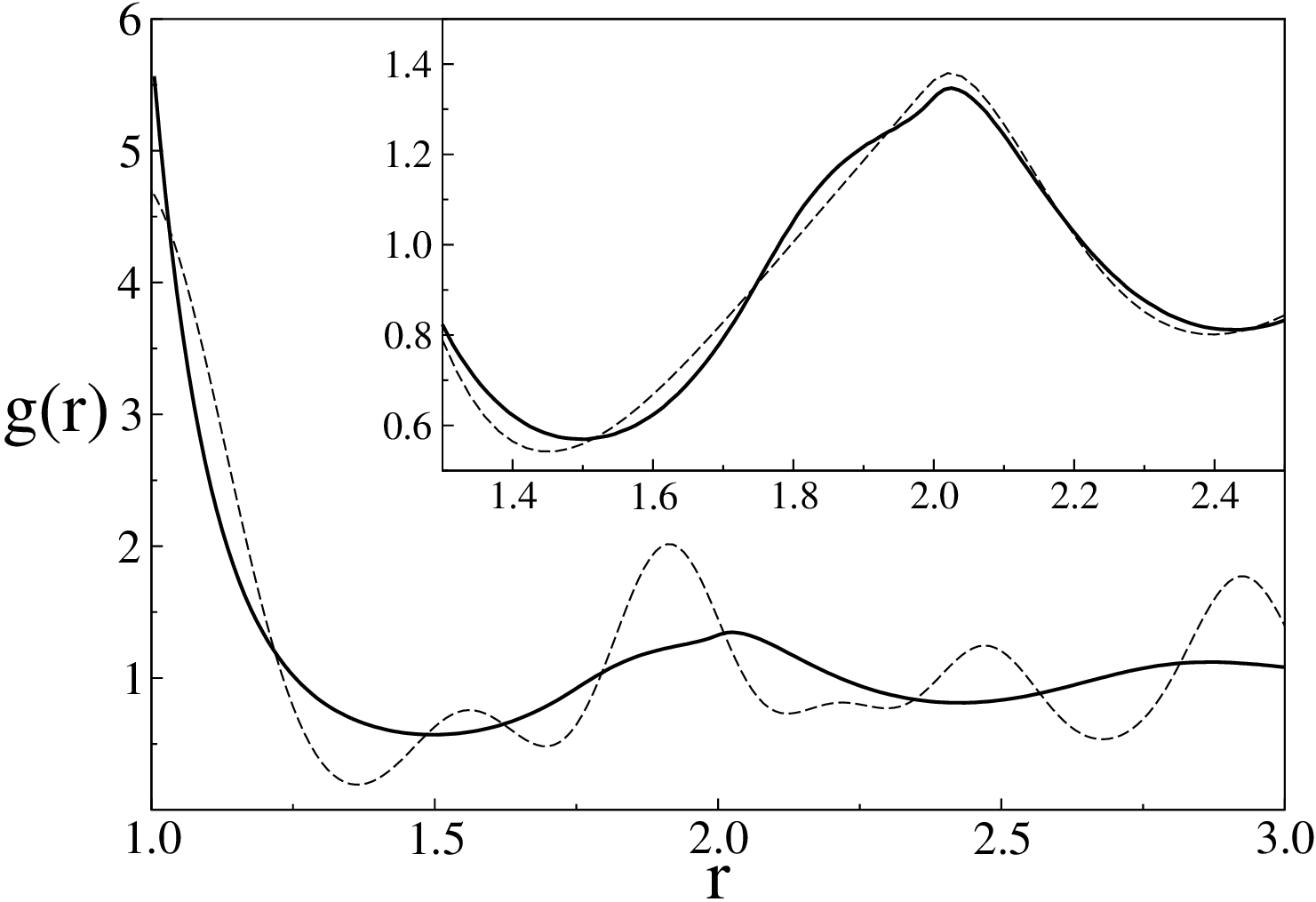}
\caption{}
\label{fig1}
\end{figure}

{\label{theoretical}}
The simulation results of Truskett {\em et al\,} \cite{truskett} show that in both two and 
three-dimensions the shoulder in the simulation RDF first becomes apparent 
within approximately five percent of the freezing transition.
For the purpose of the present work we restrict ourselves to the case of three-dimensional 
hard-spheres for which the relevant density range is 
$0.47<\eta<0.494$. The packing fraction is given by $\eta=\pi\rho\sigma^3/6$ 
for given number density $\rho$ and  
we take the sphere diameter $\sigma$ as the unit of length. 
In Figure.\ref{fig1} we show the RDF obtained from Monte-Carlo simulation for 
$\eta=0.494$ \cite{erik}, where the shoulder is most distict, together with that of the 
hard-sphere solid at the melting point, located at $\eta=0.545$ 
\cite{kincaid_weis}.
Within the high density fluid there exist ordered crystalline domains with a local 
structure similar to that of the solid phase at the melting point and which   
lend a significant statistical weight to the RDF \cite{truskett}.  
Comparison of the coexisting fluid and solid structure in Figure.\ref{fig1} 
is indeed suggestive of local crystalline ordering. 
The fluid phase RDF is enhanced (suppressed) around the peaks (troughs) of the 
coexisting solid RDF relative to what would be expected on the basis of a simple 
extrapolation of the intermediate density fluid RDF up to freezing.
The inset to Figure.\ref{fig1} shows more clearly the detailed structure of the 
second peak at freezing, in particular the shoulder which approximately spans 
the range $1.73<r<1.95$.
\par
The most commonly used theories of the pair structure of classical fluids in 
equilibrium are based upon closures of the Ornstein-Zernike equation \cite{hansen}
\begin{eqnarray}
h(r_{12})=c(r_{12}) + n\!\int \!d\rr_3 \,c(r_{13})\,h(r_{32}),
\label{oz}
\end{eqnarray}
where $n$ is the bulk density and $h(r)=g(r)-1$. Eq.(\ref{oz}) serves to define the 
direct correlation function $c(r)$ which is a function of simpler structure and 
shorter range (for finite range potentials) than $h(r)$. 
When supplemented by an independent
relation between $h(r)$ and $c(r)$, containing the details of the interaction 
potential, (\ref{oz}) yields a closed integral equation for the pair correlations 
in the system. Integral equation theories are non-perturbative in character and, 
although the approximations involved are largely uncontrolled, can produce excellent 
results for some model fluids.
The classic closure approximations are the Percus-Yevick (PY)
\cite{py} and Hyper-Netted-Chain (HNC) \cite{hnc},  
both of which can be systematically derived using diagrammatic techniques 
\cite{stell_diagram_rev}.  
For hard-spheres these theories have been subsequently improved upon by  
closures such as the Generalized-Mean-Spherical-Approximation (GMSA) \cite{gmsa}, 
Martynov-Sarkisov \cite{martynov_sarkisov},   
Ballone-Pastore-Galli-Gazzillo \cite{bpgg}, 
Verlet-Modified \cite{verlet_80}, 
Rogers-Young \cite{rogers_young} 
and Modified-Hyper-Netted-Chain \cite{mhnc}. 
Although these improved theories represent the state of the art in the theory of 
fluids, none of them captures the shoulder in the RDF close to freezing \cite{foot1}. 
In the inset to Figure.\ref{fig1} we show a representative example of this failure by 
comparing the GMSA theory with simulation at freezing. 
The triangular structure of the second peak from the GMSA theory is typical of 
integral equation theories based on (\ref{oz}), missing both the detailed structure 
of the second peak and incorrectly predicting both the location and depth 
of the first minimum. For $\eta<0.45$ all of the above mentioned theories are  
reliable and agree with simulation data to an acceptable level of accuracy. 
\par
The primary advance of the theories described above is to improve the 
thermodynamic output with respect to the PY and HNC closures by treating more 
accurately the RDF in the vicinity of contact. 
This is typically achieved by incorporating an additional parameter into the 
theory which can be determined by requirements of thermodynamic consistency.
At statepoints removed from freezing this appears to be quite adequate and leads 
to a good description of the pair correlations at all seperations. 
In order to capture the high density shoulder a treatment which describes more accurately 
local packing effects is required.
A natural extension of approaches based on (\ref{oz}) is to use instead the 
inhomogeneous Ornstein-Zernike equation \cite{bob_review,hansen} as a starting point 
and consider closure relations between the inhomogeneous direct and total correlation 
functions, $c(\rr_1,\rr_2)$ and $h(\rr_1,\rr_2)$. The inhomogeneous 
Ornstein-Zernike equation is given by
\begin{eqnarray}
h(\rr_1,\rr_2)=c(\rr_1,\rr_2) + \!\int \!d\rr_3\,c(\rr_1,\rr_3)\,n(\rr_3)h(\rr_3,\rr_2),
\label{ioz}
\end{eqnarray}
where $n(\rr)$ is the spatially varying density in the presence of an arbitrary 
external field.
In the absence of an external potential the system is translationally invariant, 
$c(\rr_1,\rr_2)\rightarrow c(r_{12})$ and 
$h(\rr_1,\rr_2)\rightarrow h(r_{12})$, 
and we recover (\ref{oz}). 
As the inhomogeneous one and two-body distribution functions exhibit the same 
symmetries as the external potential it is possible to simplify (\ref{ioz}) 
using integral transforms for the special cases of 
an external potential possessing either planar or spherical symmetry. 
In planar geometry Hankel transformation reduces (\ref{ioz}) to a one-dimensional 
integral \cite{planar_ioz_factorization}. 
This simplification has enabled application of inhomogeneous Ornstein-Zernike 
theory to both adsorption at planar substrates and the liquid-vapour interface 
\cite{planar_ioz_applications}. 
In spherical geometry the factorization of (\ref{ioz}) is achieved using 
Legendre transforms \cite{attard_py}. 
An arbitrary spherically inhomogeneous function $f(\rr_1,\rr_2)$ can be expressed 
as a function of two scalar radial variables, $r_1$ and $r_2$, and the cosine 
of the angle between them, $x_{12}\equiv \cos(\theta_{12})$.  
The Legendre transform is thus given by 
\begin{eqnarray} 
\hat{f}_n(r_1,r_2)=\frac{2n+1}{2}\int_{-1}^1 \!\!\!dx_{12}\, f(r_1,r_2,x_{12})P_n(x_{12}),
\label{legendre_transform}
\end{eqnarray}
where $P_n(x)$ is the Legendre polynomial of degree $n$. 
The inverse transform is given by
\begin{eqnarray}
f(r_1,r_2,x_{12})=\sum_{n=0}^{\infty}\hat{f}_n(r_1,r_2)P_n(x_{12}).
\label{legendre_resolution}
\end{eqnarray} 
The practical advantage of using Legendre transforms for such problems is that the 
rapid convergence of the series (\ref{legendre_resolution}) allows numerical 
calculations to be performed using a finite number of polynomials.
Applying the Legendre transformation to (\ref{ioz}) yields  \cite{attard_py}
\begin{eqnarray}
\hat{\gamma}_n(r_1,r_2) \!=\!
\frac{4\pi}{2n+1}\int_0^{\infty}\!\!\!\!\!dr_3\,r_3^2\, \hat{c}_n(r_1,r_3)
n(r_3) 
\hat{h}_n(r_3,r_2),
\label{ioz_transform}
\end{eqnarray}
where $\hat{\gamma}_n(r_1,r_2)=\hat{h}_n(r_1,r_2)-\hat{c}_n(r_1,r_2)$ has been 
introduced for notational convenience. 
As for planar geometry the spherically inhomogeneous Ornstein-Zernike equation 
(\ref{ioz_transform}) retains a one-dimensional integral. 
The transformed equation (\ref{ioz_transform}) has been used to study the structural 
properties of a number of model fluids under conditions of spherical inhomogeneity 
\cite{spherical_ioz_applications}.
\par
A special case of particular interest arises when the external potential is generated 
by a particle of the fluid fixed at the coordinate origin. In this test-particle 
limit the one-body density $n(r)$ is related to the bulk RDF by the Percus 
identity \cite{percus_identity}, $n(r)=ng(r)$, and a self-consistent theory for the 
pair and triplet correlations can be constructed.
In order to provide a closed theory the spherically inhomogeneous Ornstein-Zernike 
equation (\ref{ioz_transform}) must be supplemented 
by an equation for the one-body density $n(r)$ and a closure relation between 
$c(r_1,r_2,x_{12})$ and $h(r_1,r_2,x_{12})$. 
Several exact relations exist which express $n(r)$ as a functional of the 
inhomogeneous pair correlation functions. 
Of particular importance are the first member of the Yvon-Born-Green (YBG) hierarchy 
\cite{ybg}, the Triezenberg-Zwanzig (TZ) equation \cite{triezenberg_zwanzig} and the 
Lovett-Mou-Buff-Wertheim (LMBW) equation \cite{lmbw}. 
We choose to work with the LMBW equation 
\begin{eqnarray}
\nabla n(\rr_1) \!= \!-n(\rr_1)\left( \nabla V(\rr_1) 
+ \int \!\!d\rr_2\, c(\rr_1,\rr_2)\nabla n(\rr_2)\right),
\label{lmbw}
\end{eqnarray}
where $V(\rr_1)$ is the external potential in units of $k_BT$ \cite{rushbrooke}. 
In the test-particle limit the external potential is 
equal to the interparticle pair potential, $V(\rr)=\phi(r)$, and the LMBW 
equation reduces to  
\begin{eqnarray}
\frac{n'(r_1)}{n(r_1)} =-\phi'(r_1) 
+ \frac{4\pi}{3}\!\!\int_0^{\infty} \!\!\!\!dr_2\,r_2^2\, \hat{c}_{\,1}(r_1,r_2)\,n'(r_2), 
\label{lmbw_spherical}
\end{eqnarray} 
where $\hat{c}_{\,1}(r_1,r_2)$ enters as a result of the angular integrals.
The YBG,TZ and LMBW equations all yield exact relations between $n(r)$ and $\phi(r)$, 
given the exact inhomogeneous correlation functions as input.  
If $c(\rr_1,\rr_2)$  and 
$h(\rr_1,\rr_2)$ are only known approximately, as is always the case in practice, 
then each of the three equations will yield a different $n(r)$ for a 
given $\phi(r)$. This structural inconsistency is analogous to the 
thermodynamic inconsistency familiar from standard integral equation 
theories based on (\ref{oz}), where each of the three thermodynamic routes 
(virial, compressibility and energy) yields a different result for the pressure. 
That an additional inconsistency in the pair correlations occurs between the 
YBG,TZ and LMBW equations is not surprising and simply reflects that 
we are working at a higher level in the hierarchy. 
Our choice to use the LMBW equation is motivated by the fact that for many 
situations the derivative $n'(r)$ decays more rapidly than $n(r)$. 
This is advantageous for numerical implementation as
the integral can be discretized over a smaller range. 
\par
The exact equations (\ref{ioz_transform}) and (\ref{lmbw_spherical}) need to be 
supplemented by a closure relation between $c(\rr_1,\rr_2)$ and $h(\rr_1,\rr_2)$.  
This closure represents the only approximation within our treatment. 
Experience with approximate closures of the homogeneous Ornstein-Zernike equation 
(\ref{oz}) has shown that for statepoints removed from freezing 
a suitably chosen {\em local} closure relation between the direct and total correlation 
function can produce reliable results for the pair structure. 
We adopt here the same strategy and consider applying the same local closure relations 
between the inhomogeneous functions as are usually applied between the corresponding 
homogeneous functions.  
In the case of the PY and HNC equations such a generalization is natural as 
both theories follow from well defined diagrammatic resummations. 
The resummation procedure is not altered by the association of field points in 
the Mayer cluster diagrams with the spatially varying one-body density $n(\rr)$ 
\cite{stell_diagram_rev}. 
Nevertheless, it is not clear that the fortuitous cancellation of errors which can 
occur when applying a given closure to (\ref{oz}) (e.g. hard-spheres in the PY 
approximation) will be retained when applying the same closure to (\ref{ioz}).
Moreover, application of thermodynamically consistent closures 
at the triplet level does not guarantee improved results,   
particularly for the description of structure in regions close to the source of 
inhomogeneity. 
Despite these potential shortcomings the ultimate assessment of the validity of 
this approach can only be made from the numerical results, and these are currently 
very encouraging. 
Studies of simple model fluids in planar \cite{planar_ioz_applications} and spherical 
\cite{spherical_ioz_applications} geometry suggest that significant improvement over 
standard (homogeneous) integral equation theory can be achieved.  
\par
In spherical geometry the formally exact closure relation is given by 
\begin{eqnarray}
h(r_1,r_2,x_{12})+1\!&=&\!\exp[-\phi(r_{12})\!+\!h(r_1,r_2,x_{12})
\notag\\
\!\!&-&\!\!c(r_1,r_2,x_{12})\!+\!b(r_1,r_2,x_{12})\,],
\label{bridge}
\end{eqnarray}
where $b(r_1,r_2,x_{12})$ is the bridge function, an intractable sum of elementary 
diagrams \cite{hansen}. 
From among the various improved theories an accurate and computationally efficient approximation 
for the bridge function is provided by the spherically inhomogeneous generalization of the 
Verlet-Modified closure \cite{choudhury1}
\begin{eqnarray}
b(r_1,r_2,x_{12})) = -\frac{1}{2}\frac{\tau^2(r_1,r_2,x_{12})}
{(\,1 + \alpha\max[\tau(r_1,r_2,x_{12}),0]\,)},
\label{verlet_mod}
\end{eqnarray}
where the function $\max[\,\cdot\, , \,\cdot\,]$ returns the largest of the two 
arguments and the renormalized indirect correlation function is given by
\begin{eqnarray}
\tau(r_1,r_2,x_{12})\!=\!h(r_1,r_2,x_{12})\!-\!c(r_1,r_2,x_{12})\!-\!\phi_{\rm att}(r_{12}),
\notag\\
\label{indirect}
\end{eqnarray}
where $\phi_{\rm att}(r)$ is the attractive part of the pair potential in units 
of $k_BT$.
When (\ref{verlet_mod}) is applied as a closure to (\ref{oz}) for the hard-sphere 
system the parameter $\alpha$ is usually taken to have the value $0.8$ 
\cite{verlet_80,choudhury}.
The motivation for this choice 
is that the thermodynamic inconsistency between the compressibility and virial routes 
to the pressure remains small (within approximately two percent \cite{verlet_80}) at 
all fluid packing fractions. 
While neglecting possible density dependence of $\alpha$ only yields approximate 
thermodynamic consistency, it provides the computational advantage that the integral 
equation has only to be solved once for a given statepoint. 
The numerical demands of solving (\ref{ioz}) make this an essential requirement for 
practical application of the inhomogeneous integral equation method. 
When (\ref{verlet_mod}) is applied as a closure to the inhomogeneous Ornstein-Zernike 
equation (\ref{ioz}) the value $\alpha=0.8$ is no longer optimal for global 
thermodynamic consistency. 
For the hard-sphere system we find $\alpha=0.95$ to be a better choice 
(see the discussion in Section \ref{hardspheres} below) and it is this value which is 
employed in all numercial calculations presented in this work.
A full discussion of the merits of the closure (\ref{verlet_mod}) as applied to 
(\ref{oz}) can be found in \cite{choudhury}.
\par
Equations (\ref{ioz_transform}), (\ref{lmbw_spherical}), (\ref{bridge}) and 
(\ref{verlet_mod}) form a closed set which can be solved self-consistently for 
$n(r)$, $h(r_1,r_2,x_{12})$ and $c(r_1,r_2,x_{12})$ for given bulk density $n$, 
temperature and 
pair potential $\phi(r)$. 
The RDF is easily obtained from the density profile using $g(r)=n(r)/n$ 
whereas the bulk triplet-RDF is simply related to the 
spherically inhomogeneous RDF, $g(\rr_1,\rr_2)=h(\rr_1,\rr_2)-1$, by the expression
\begin{eqnarray}
g^{(3)}(r_1,r_2,r_{12})=g(r_1)g(r_2)g_0(\rr_1,\rr_2).
\label{triplet}
\end{eqnarray}
We employ the notation $g_0(\rr_1,\rr_2)$ to indicate that the inhomogeneity is 
due to a test-particle at the origin. 
Eq.(\ref{triplet}) is essentially the extension of the Percus identity to 
the next level in the hierarchy of correlation functions. 
Within this framework the familiar Kirkwood superposition approximation 
\cite{kirkwood} corresponds to 
$g_0(\rr_1,\rr_2)\thickapprox g(r_{12})$. 
\par
The set of equations (\ref{ioz_transform}), (\ref{lmbw_spherical}), (\ref{bridge}) and 
(\ref{verlet_mod}) were solved self consistently by standard iteration, applying 
Broyles mixing \cite{hansen} to both the inhomogeneous pair correlation 
functions and the density profile. 
A mixing parameter between $0.05$ and $0.1$ was generally found to be sufficient to 
achieve steady convergence. 
The convergence rate can be improved by employing the mixing scheme of 
Ng \cite{Ng}, but we find that the slight reduction in the nember of iterations 
does not compensate for the additional memory requirements associated 
with this method. 
The majority of computation time in numerical solution of the inhomogeneous equations 
is taken by calculation of the Legendre transforms (\ref{legendre_transform}), 
(\ref{legendre_resolution}) and the integral (\ref{ioz_transform}). 
In order to minimize computation time we have employed the discrete orthogonal 
Legendre transform proposed by Attard \cite{attard_py} which is both efficient 
and prevents the build up of round-off errors during iteration. 
The mesh for the angular integration (\ref{legendre_transform}) is defined by the 
roots of $P_n(x)$, where $n$ is the highest-order polynomial considered. 
A cut-off of $8\sigma$ was found to be sufficient for all calculations presented 
in the present work. 
Corrections for the long range tails were made following the suggestions of 
\cite{attard_tails, fushiki} whereby the inhomogeneous functions are replaced by their 
bulk counterparts far from the source of inhomogeneity.
In order to obtain high accuracy solutions we employ a grid spacing $dr=0.02$ and 
between $100$ and $140$ Legendre polynomials in the sum (\ref{legendre_resolution}). 
For hard spheres at packing fractions $\eta<0.4$ accurate solutions can be obtained 
with a coarser mesh.   
At high density statepoints, for which the liquid is highly structured, it is necessary
to work with such a large number of polynomials and a fine grid in order to ensure 
convergence.
For further details of the numerical method we refer the reader to 
the original work of Attard \cite{attard_py}.

%
\section{Results}{\label{results}}
\subsection{Hard spheres}{\label{hardspheres}}
We now consider application of equations (\ref{ioz_transform}), 
(\ref{lmbw_spherical}), (\ref{bridge}) and (\ref{verlet_mod}) to the hard-sphere 
system at packing fractions close to freezing. 
In the spirit of the the Verlet-Modified approximation \cite{verlet_80} we have 
determined the value of the parameter $\alpha$ in (\ref{verlet_mod}) by 
requiring that the pressures obtained from the virial and compressibility equations 
agree as closely as possible 
over the entire fluid density range. This is a numerically demanding procedure 
which requires solution of the integral equation on an $\eta$-grid covering the 
range $0<\eta<0.494$ for each trial value of $\alpha$. 
Numerical accuracy is an issue when calculating the pressure, particularly via the 
compressibility route, and places limits on the accuracy to which $\alpha$ can be 
tuned. We find that the value $\alpha=0.95$ effectively minimizes the difference 
in the pressures obtained from the two routes, with a residual discrepancy around 
the one percent level. 
\par
\begin{figure}
\newpage
\hspace*{-0.5cm}
\includegraphics[width=10cm,angle=0]{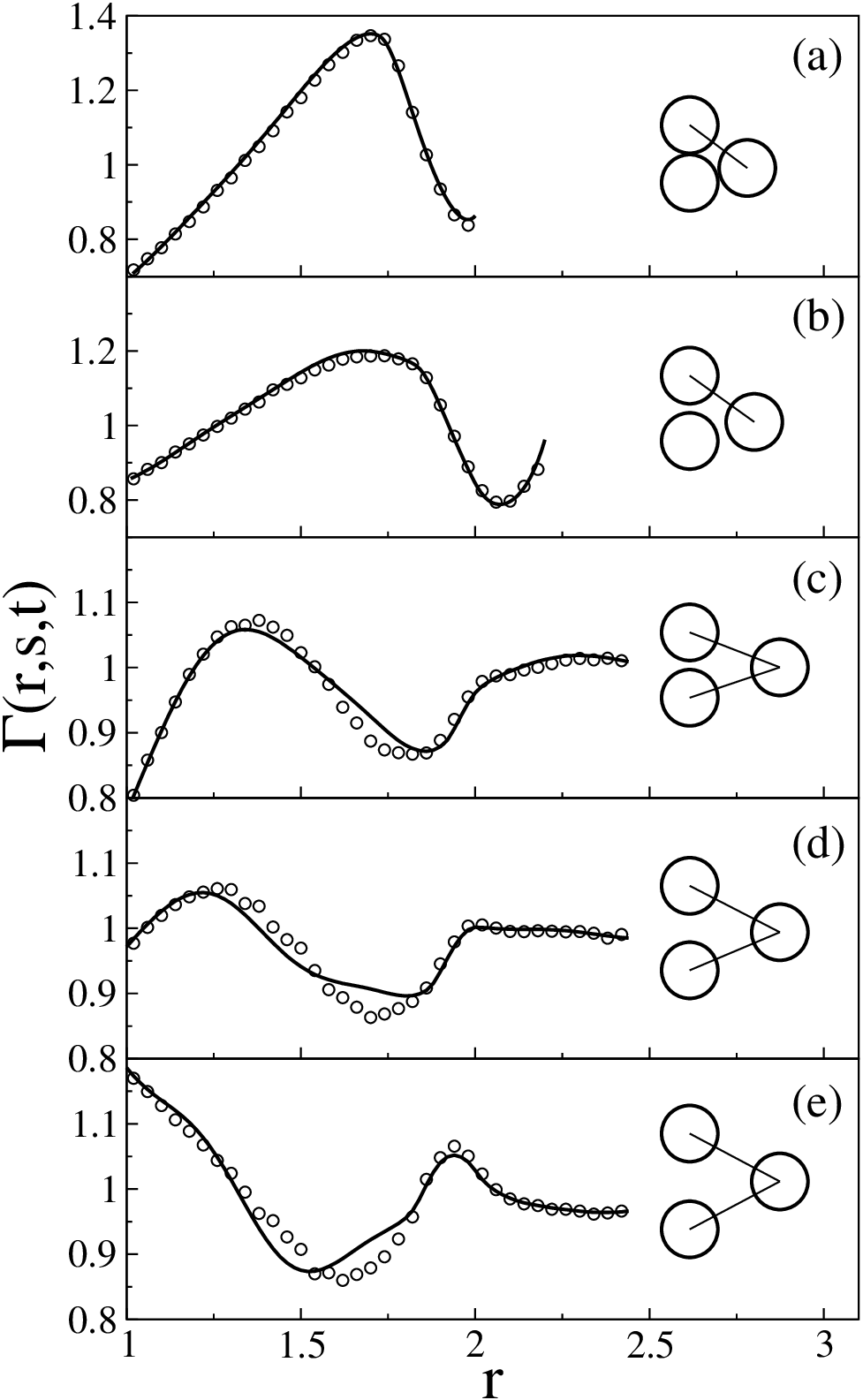}
\caption{}
\label{fig2}
\end{figure}

Before working close to the freezing transition we first seek to establish the 
accuracy of our approach at dense fluid statepoints removed from freezing. 
Monte-Carlo simulations have been performed for the pair and triplet 
correlations of the hard sphere system at $\eta=0.4189$ and serve as a useful 
benchmark for our theory \cite{mueller_gubbins}. 
We find that 
at this packing fraction the RDF from the inhomogeneous integral equation theory 
lies perfectly within the error bars of the simulation data (not shown). 
As the theory is apparently quasi-exact for the pair correlations we omit this 
comparison and focus instead 
on the triplet-RDF for which more significant deviations can be detected. 
In Figure.\ref{fig2} we show results for the quantity 
\begin{eqnarray}
\Gamma(r,s,t)=\frac{g^{(3)}(r,s,t)}{g(r)g(s)g(t)},
\label{big_gamma}
\end{eqnarray}
for a selection of specific configurations. 
The quantity $\Gamma(r,s,t)$ measures deviations from the superposition approximation 
and is a sensitive indicator of errors in theoretical approximations for the 
triplet-RDF. 
Figs.\ref{fig2}a and \ref{fig2}b show $\Gamma(s,s,r)$ as a function of $r$ for 
two different values of $s$ (rolling geometries). 
Given the high density, the level of agreement between theory and simulation is 
excellent and essentially all details of the simulation data are faithfully reproduced.
It should be noted that use of the test-particle approach in combination with an 
approximate closure such as (\ref{verlet_mod}) leads to a $g^{(3)}(r,s,t)$ which 
does not respect the exact particle exchange symmetry expected of this function.
We have investigated this discrepancy for each of the configurations shown in 
Fig.\ref{fig2} and find that the curves differ by at most a few percent, depending 
upon the location of the test particle in the triangle. 
This close agreement validates to some extent use of the closure (\ref{verlet_mod}) 
as the approximate particle exchange symmetry is a non-trivial output of the 
theory. 
The data shown in Fig.\ref{fig2} were generated with the test particle at the lower 
left position in the schematic configurations shown. 
In Figs.\ref{fig2}c-\ref{fig2}e we consider isosceles triangle configurations for 
three different triangle base lengths. 
For $0<r<1.4$ and $r>1.8$ the simulation data are well described by the theory.  
In particular, the increase of the contact value and development of a peak at 
$r=1.9$ in going from \ref{fig2}d to \ref{fig2}e are correctly captured.
In the range $1.4<r<1.8$ there are slight discrepancies between theory and 
simulation as the first minimum is not correctly located.
It is reassuring to note that the self-consistently obtained RDF appears 
insensitive to these fine details and remains accurate, despite 
modest errors in the underlying triplet structure.
We can thus proceed with confidence to perform calculations 
closer to freezing.  


%
\par
\begin{figure}
\newpage
\hspace{-0.5cm}
\includegraphics[width=14cm,angle=0]{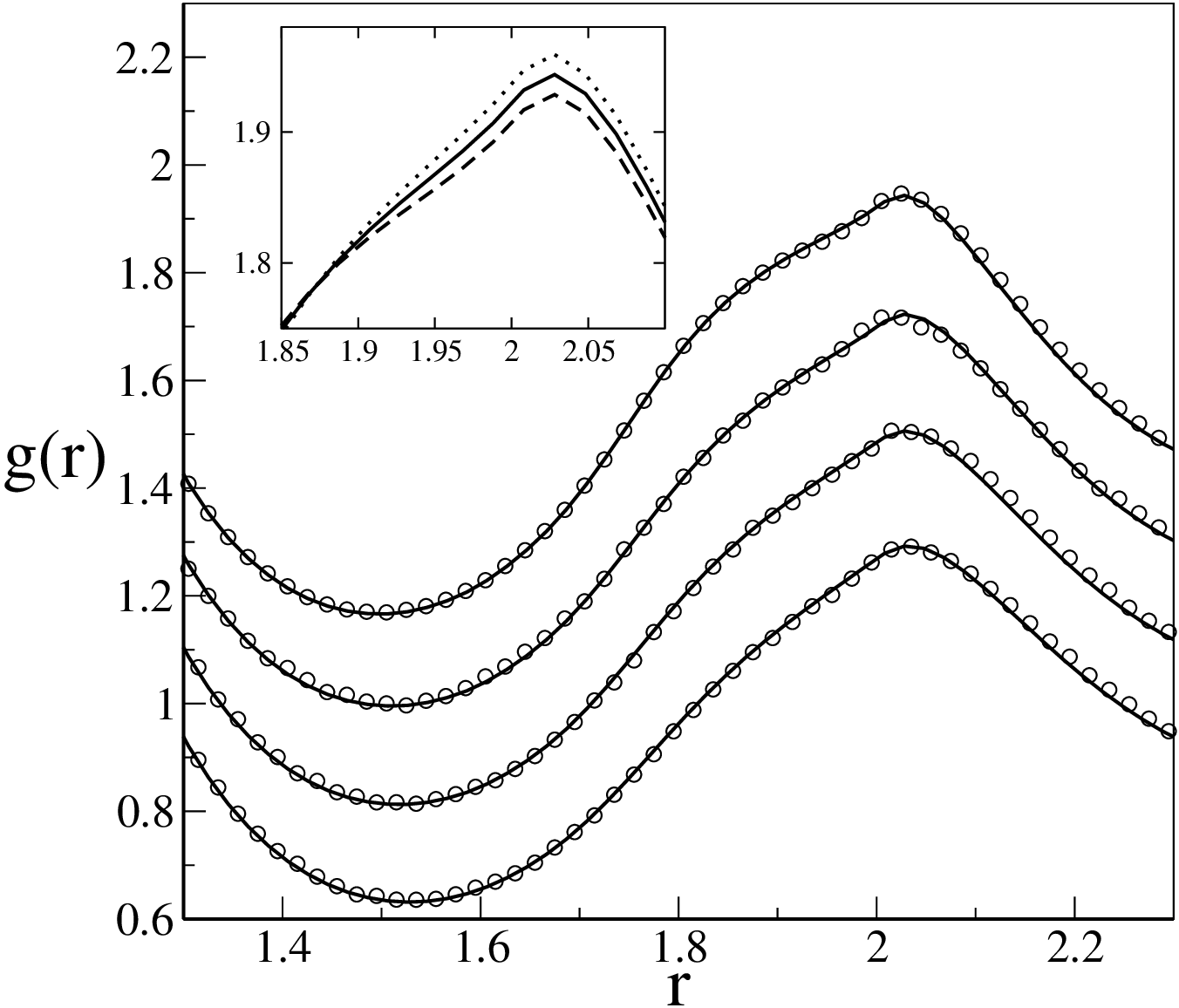}
\caption{}
\label{fig3}
\end{figure}

In Fig.\ref{fig3} we show results for the RDF at packing fractions 
$\eta=0.46, 0.47, 0.48$  and $0.494$ and compare with the results of molecular dynamics 
simulations \cite{erik}.  
We concentrate on the range $1.3\le r\le 2.4$ in order to focus attention on the 
second peak.  
At packing fraction $\eta=0.46$ the shoulder is just visible in 
the simulation RDF, marking the onset of local crystalline ordering in the fluid. 
Remarkably, this behaviour is correctly captured by the present inhomogeneous 
integral equation theory.  
Both the shape of the first minimum and the point of inflection which occurs between the 
first minimum and the second peak are accurately reproduced. 
As the density is increased the shoulder develops further and by 
$\eta=0.494$ is very pronounced.  
The present theory describes almost perfectly the evolution of this rich
second peak structure as a function of density and captures the 
characteristic shape of the simulation curves, quite distinct from the 
triangular form typical of standard approaches based on (\ref{oz}).
To the best of our knowledge these findings provide the first convincing evidence 
that the method of integral equations can indicate the existence of the freezing 
transition. 
Moreover, the set of equations (\ref{ioz_transform}), 
(\ref{lmbw_spherical}), (\ref{bridge}) and (\ref{verlet_mod}) represent the first 
example of an approximate Ornstein-Zernike closure capable of reproducing from first 
principles the RDF shoulder.
In order to demonstrate the relative insensitivity of our results to changes 
in the parameter $\alpha$ we show in the inset to Fig.\ref{fig3} results for values 
above and below the optimal value $\alpha=0.95$.
\par
\begin{figure}
\newpage
\hspace{-0.5cm}
\includegraphics[width=14cm,angle=0]{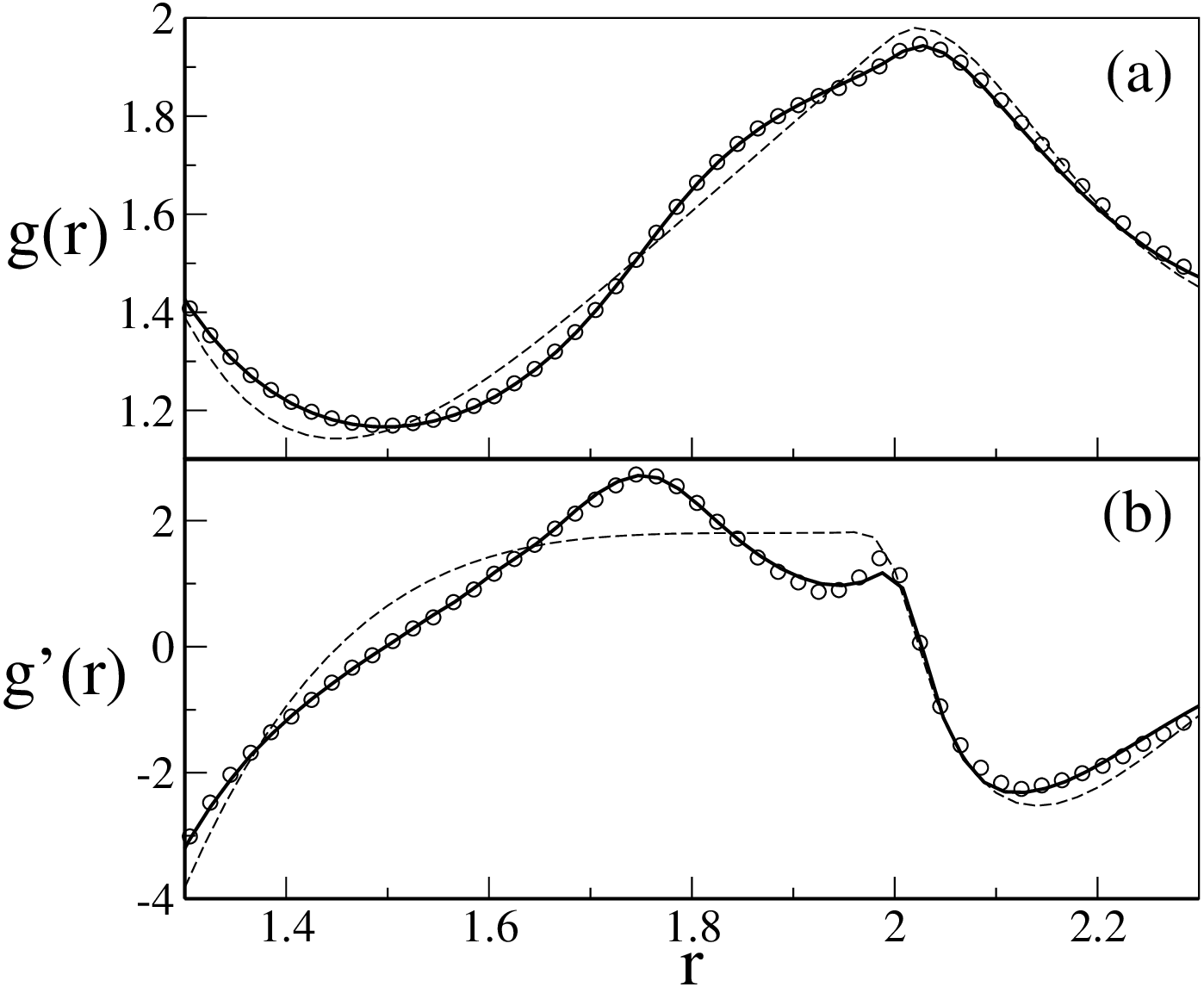}
\caption{}
\label{fig4}
\end{figure}

Fig.\ref{fig4}a and Fig.\ref{fig4}b allow a more detailed comparison to be made between 
theory and simulation, showing the RDF and its derivative at freezing.  
In Fig.\ref{fig4}a we compare the present theory with the result of both simulation and 
GMSA theory \cite{gmsa} in order to emphasize the inability of such standard integral equation 
approaches to follow the detailed structure of the simulation curves. 
In Fig.\ref{fig4}b we compare the first derivative $g'(r)$ from the present theory with 
simulation and the GMSA. 
This comparison shows clearly the level of accuracy provided by the inhomogeneous 
theory. 
The only slight deviation of the theory from the simulation data occurs in the 
region in the immediate vicinity of the second peak. 
This should be contrasted with the GMSA theory which predicts an approximately constant 
slope over the range $1.6<r<2$.     
In addition to the accurate description of the second peak we also find that the first 
peak and contact value of the RDF are in close agreement with simulation. 
The resulting virial pressure agrees very favourably with established results for the 
hard-sphere equation-of-state \cite{malijevsky_eos,hs_eos}. 

\subsection{Short-range attraction}{\label{attraction}}
The simulation studies \cite{truskett,snook} discussed in the introduction provide 
strong evidence connecting the appearance of a shoulder in the RDF of hard-spheres 
and hard-disks to the freezing transition. 
In the previous section we have shown that for hard-spheres the inhomogeneous 
integral equation theory yields an RDF consistent with this picture. 
It is therefore tempting to conclude that the appearance of a shoulder in the 
theoretical RDF can be used as an empirical indicator for locating the freezing 
transition.  
However, there remains the possibility that the occurrance of the shoulder and its
apparent connection to the onset of freezing may be particular to the hard-sphere 
(hard-disk) system. 
We thus seek to investigate the generality of this connection by applying the 
inhomogeneous theory to a system interacting via a hard-sphere repulsion plus a 
short-range attraction.
For such interaction potentials it is well known that when the range of 
the attractive interaction becomes sufficiently short-ranged the critical point 
of the liquid vapour transition becomes metastable with respect to freezing. 
This is a result of the rapid broadening of the fluid-solid coexistence region as a 
function of attraction strength which overlaps entirely with the liquid-vapour 
coexistence region. 
This novel phase diagram topology thus makes it possible to approach the 
freezing phase boundary by increasing attraction strength at fixed density. 
The confirmation of a shoulder in the RDF in the vicinity of this more general 
freezing phase boundary would considerably strengthen the argument for this empirical 
freezing indicator. 
\par
A well studied pair-potential yielding the required phase diagram topology is that due 
to Asakura, Oosawa and Vrij \cite{asakura}. 
The potential consists of a hard-sphere repulsion plus an attractive tail given by 
\begin{eqnarray}
\phi_{\rm AO}(r)=\eta^r_p\frac{(1+q)^3}{q^3}
\left(
1 - \frac{3r}{2(1+q)} + \frac{r^3}{2(1+q)^3}
\right)
\label{ao}
\end{eqnarray}
for $1<r<1+q$ and zero otherwise. 
The parameter $q$ sets the range of the potential and the amplitude $\eta^r_p$ 
determines the depth of the potential well at contact. 
This potential has stimulated much interest as it provides a simple approximation 
to the depletion potential between two hard-sphere colloids in a suspension with 
added non-adsorbing polymer \cite{gast,dijkstra}. 
In this context $\eta^r_p$ is the packing fraction of polymer coils in a reservoir 
attached to the system. 
For the purpose of the present work $\eta^r_p$ may simply be regarded as a parameter 
determining the strength of the attractive interaction. 
The potential (\ref{ao}) is convenient because there exist simulation 
data for the phase boundaries in the ($\eta_p^r,\eta_c$) plane for several values of 
the range parameter $q$. 
In Fig.\ref{fig5} we show the simulation phase diagram for $q=0.4$ \cite{dijkstra}. 
For this value of $q$ the liquid-vapour phase boundary is metastable, albeit weakly, 
with respect to freezing and the potential is only moderately short-range. 
For very short-range potentials ($q< 0.2$) 
resolution of the high and 
narrow first peak in the RDF requires a spatial grid finer than the $dr=0.02$ 
presently employed and leads to unacceptable computational demands. 
The choice $q=0.4$ thus represents a reasonable compromise, being both numerically 
tractable and displaying the desired phase diagram topology.  
\par 

%
In order to analyse the structural predictions of the inhomogeneous theory for the 
potential (\ref{ao}) we approach the freezing boundary along two distict paths in 
the phase diagram. 
The statepoints for which we present detailed results are indicated in Fig.\ref{fig5}.
We consider first the packing fraction $\eta=0.35$ and investigate 
the RDF as a function of increasing $\eta_p^r$, approaching the freezing boundary along 
a vertical path in the phase diagram. 
In Fig.\ref{fig6} we show the RDF in the vicinity of the second peak for four different 
statepoints. 
At statepoint (a) the RDF is simply that of hard-spheres ($\eta_p^r=0$) and displays 
the expected oscillatory structure. 
As the attraction is increased (statepoint (b)) there is a notable  
increase in structure over that of pure hard-spheres. 
The shifting of the first minimum and second peak to smaller separations reflects the 
increased interparticle attraction which tends to pull the next-nearest-neighbours 
closer to the central particle. 
Nevertheless, the results for statepoints (a) and (b) can still be well reproduced 
by standard integral equation closures. 
At statepoint (c) additional fine structure not captured by standard
theories becomes evident in the second peak and a marginal point of inflection occurs.
For slightly stronger attraction (statepoint (d)) this structure  
develops into a distict shoulder, reminiscent of that seen close to the freezing 
transition of pure hard-spheres. 
The development of the shoulder between statepoints (c) and (d) is consistent 
with the broadened freezing transition found in simulation, as shown in Fig.\ref{fig5}, 
and occurs over a relatively narrow range of attraction strengths 
($0.43<\eta_p^r<0.47$). 
It may be inferred that the appearance of this attraction-driven shoulder reflects 
the existence of underlying precursor structures in the intermediate density fluid. 
In Fig.\ref{fig7} we consider approaching the freezing boundary at the lower packing 
fraction $\eta=0.2$.
Increasing the attraction strength from statepoint (e) to (g) we observe the expected 
increase in fluid structure. 
Statepoint (h) lies just below the freezing phase boundary and exhibits 
the onset of a weak shoulder which 
develops as the attraction strength is further increased to statepoint (i). 
In accord with the findings for $\eta=0.35$ the onset of the shoulder occurs over a 
relatively narrow range of attraction strength and is consistent with the location 
of the simulation phase boundary.
For this lower packing fraction the shoulder is less pronounced and is shifted to 
slightly smaller separations, indicating the decreased statistical weight of the 
locally ordered regions within the fluid. 
The striking correlation between the onset of the shoulder and the simulation 
phase boundary lead us to conclude that the RDF shoulder is a genuine freezing 
indicator for a broad class of model interaction potentials and not simply an 
artifact of the hard-sphere fluid at high density.     

%

\section{Discussion}{\label{discussion}}
In this paper we have demonstrated that inhomogeneous integral equation theory 
provides a highly accurate description of the hard-sphere fluid RDF at densities 
up to the freezing transition. 
In agreement with simulation the theory displays the onset of a shoulder in the 
second peak which is associated with the formation of local crystalline regions 
in the dense fluid. 
When applied to a system of hard-spheres plus a short-range attraction the 
inhomogeneous theory predicts the development of a shoulder 
as a function of attraction strength consistent with the simulation freezing 
phase boundary. 
That the inhomogeneous integral equation theory is capable of capturing these 
subtle effects demonstrates the accuracy with which local packing constraints 
are treated. 
The fact that the present theory acknowledges the existence of the freezing 
boundary is a significant development beyond standard integral equation 
closures and validates the significant increase in computational resources required 
for numerical solution of the equations. 
Although the present study is restricted to three dimensional systems 
we anticipate that application of the inhomogeneous theory in two dimensions 
would lead to broadly similar conclusions. 
\par

%
For the purpose of this work we have focussed on fluid state structure in 
the vicinity of freezing. However, the inhomogeneous theory is also capable of a 
description of the liquid-vapour phase transition which occurs for systems 
with longer-range attraction than that considered here. 
For $q>0.45$ the potential (\ref{ao}) exhibits a stable liquid-vapour transition. 
Preliminary calculations for range parameter $q=0.6$ display a large increase in 
the compressibility along a locus of points in the phase diagram consistent 
with the simulation liquid-vapour phase boundary \cite{dijkstra}. 
We therefore feel justified in claiming the theory developed in this work to 
be the first example of an Ornstein-Zernike closure sensitive to both the second-order 
liquid vapour transition and the first-order freezing transition. 
This goes a step beyond existing integral equation approaches which at best display  
a line in the phase diagram upon which the compressibility either diverges 
(spinodal) or the theory breaks down (no-solutions boundary) \cite{brader_ijtp} and 
which show no indication of freezing. 
Of particular interest would be the characterization of the spinodal
for the present theory and the determination of the critical exponents 
which, given the nature of the closure, may be non-classical \cite{parola}. 

\subsection*{Acknowledgements}
We acknowledge the transregio SFB TR6 for financial support.
We thank E. Lange for stimulating discussions and for providing 
molecular dynamics data.

\newpage
Figure 1. Hard-sphere RDF from computer 
simulation at freezing, $\eta=0.494$ (full line) and at the melting point 
$\eta=0.545$ (dashed line).
The inset focuses on the shoulder in the second peak of the simulation 
RDF at freezing (full line). 
The broken line is the result of GMSA theory and is representative of the inability 
of standard integral equation theories to capture the shoulder.\\

Figure 2. Comparison of the results of inhomogeneous integral equation theory 
with the simulation results of M\"uller and Gubbins \cite{mueller_gubbins} 
for the quantity $\Gamma(r,s,t)$ for $\eta=0.8\pi/6\thickapprox 0.4189$. 
The geometries are schematically shown in each figure. The separation $r$ is 
indicated by a line in each case. (a) and (b) are rolling geometries at 
$s=t=1.0$ and $s=t=1.1$, respectively. (c)-(e) are isoceles triangle 
configurations with $s=r$ and the base length of the triangle fixed at 
$t=1.1$, $1.3$ and $1.5$, respectively.\\

Figure 3. The RDF of hard-spheres for packing fractions $\eta=0.46, 0.47, 0.48$ 
and $0.494$ (from bottem to top). 
For clarity the data for $\eta=0.47, 0.48$ and $0.494$ 
have been shifted vertically by $0.2, 0.4$ and $0.6$, respectively.
Solid lines are the results of the inhomogeneous integral equation theory. 
Open circles are from molecular dynamics simulations \cite{erik}.
The inset shows the sensitivity of the second peak to changes in the parameter 
$\alpha$ for $\eta=0.494$. Results are shown for the optimal value $\alpha=0.95$ 
(solid line), $\alpha=1.05$ (broken line) and $\alpha=0.85$ 
(dotted line). 
\\

Figure 4. Comparison of the inhomogeneous integral equation theory (solid line) for $\eta=0.494$ 
with simulation data (open circles) \cite{erik} and the GMSA theory 
(dashed line) \cite{gmsa}. (a) and (b) show the RDF and the first derivative 
of the RDF, respectively. \\

Figure 5. The simulation phase diagram calculated using the   
Asakura-Oosawa pair potential (\ref{ao}) in the $(\eta,\eta^r_p)$ 
plane \cite{dijkstra}. 
The range of attraction is $q=0.4$ and $\eta^r_p$ represents the 
strength of the attraction. The solid line is the fluid-solid phase boundary 
and the dashed line indicates the (metastable) liquid-vapour transition. 
Squares labelled a-d and circles labelled e-i 
mark the state points for which we present detailed results.\\

\begin{figure}
\newpage
\vspace{0.0cm}
\hspace{-0.1cm}
\includegraphics[width=14cm,angle=0]{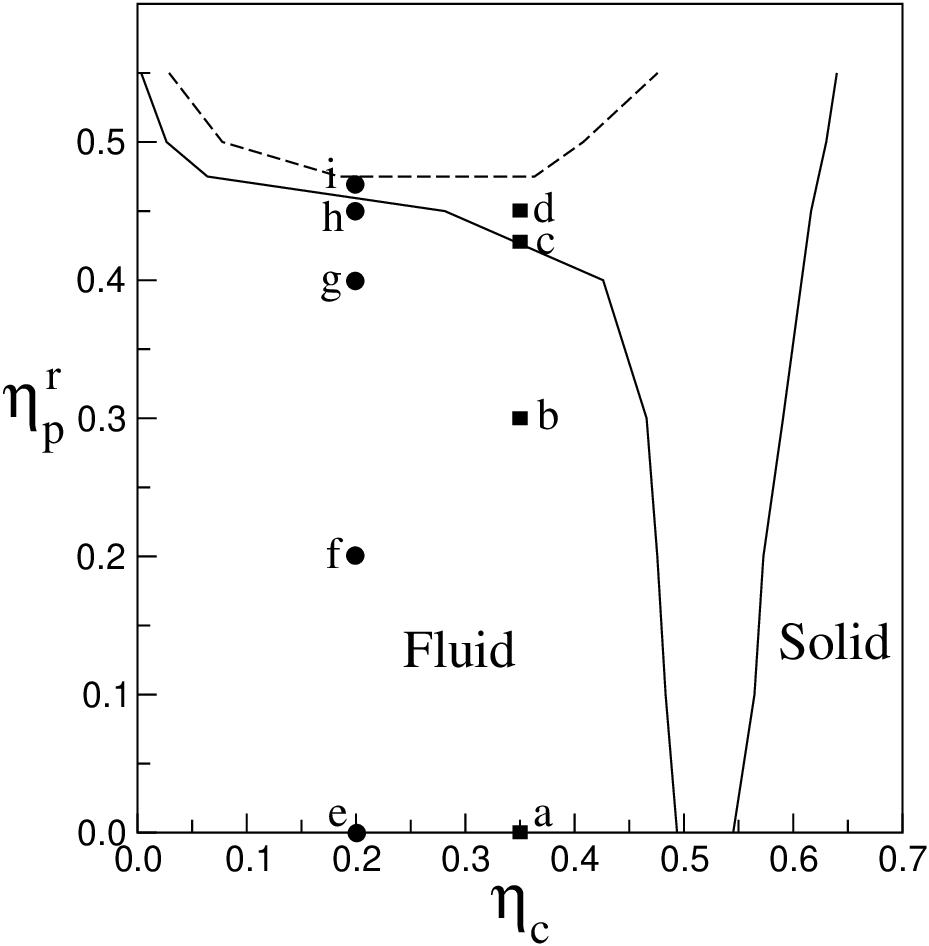}
\caption{}
\label{fig5}
\end{figure}

Figure 6. The RDF from inhomogeneous integral equation theory for the 
statepoints labelled a-d in Fig.\ref{fig5}. In each case the packing fraction  
$\eta=0.35$. Statepoint (a) $\eta_p^r=0$ (dotted line), (b) $\eta_p^r=0.3$ 
(dot-dashed line), (c) $\eta_p^r=0.43$ (broken line) and (d) 
$\eta_p^r=0.47$ (full line). The development of the shoulder is consistent 
with the simulation freezing phase boundary presented in Fig.\ref{fig5}.\\

\begin{figure}
\newpage
\vspace{0.0cm}
\hspace{-0.5cm}
\includegraphics[width=14cm,angle=0]{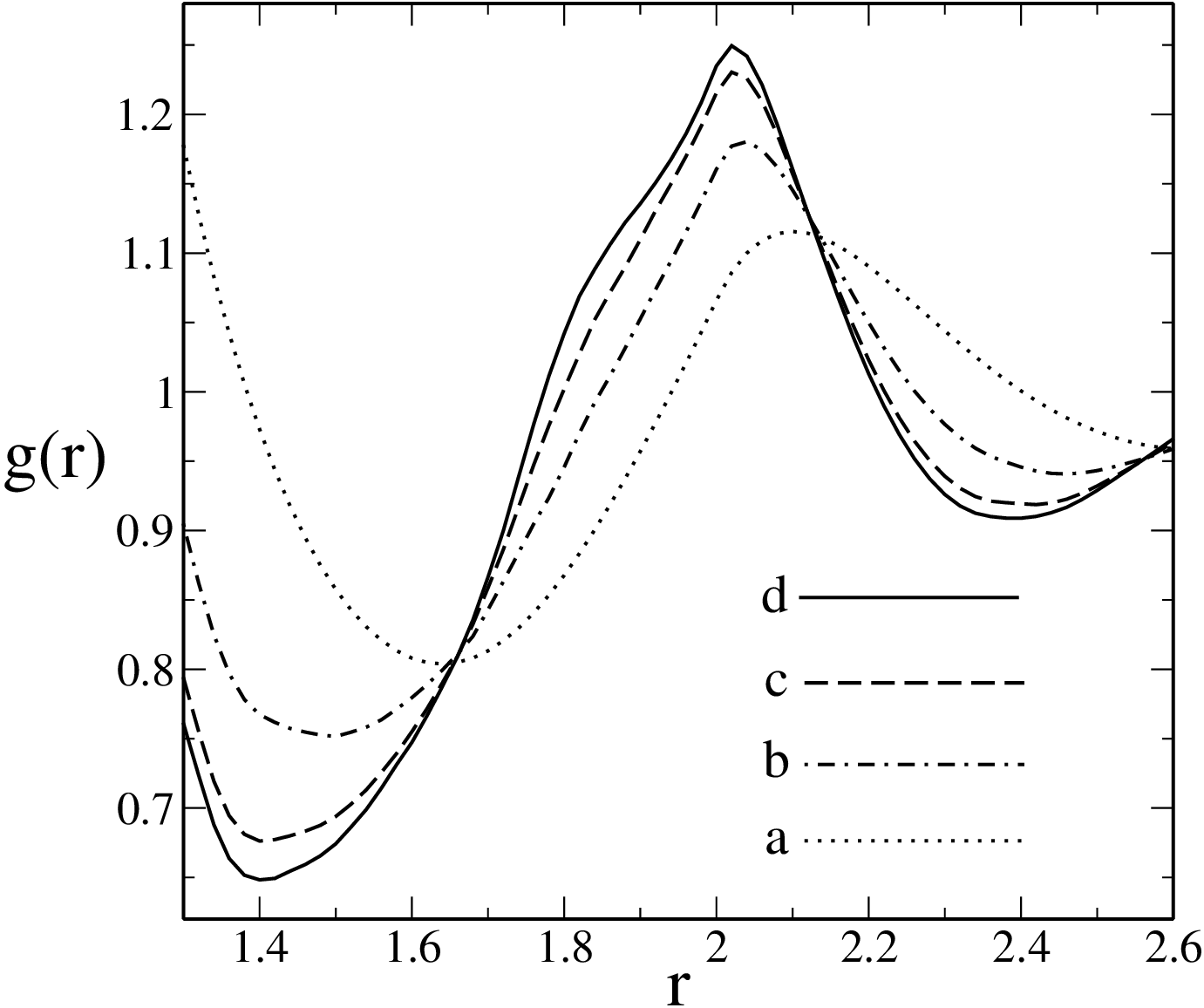}
\caption{}
\label{fig6}
\end{figure}

\begin{figure}
\newpage
\vspace{0.0cm}
\hspace{-0.5cm}
\includegraphics[width=14cm,angle=0]{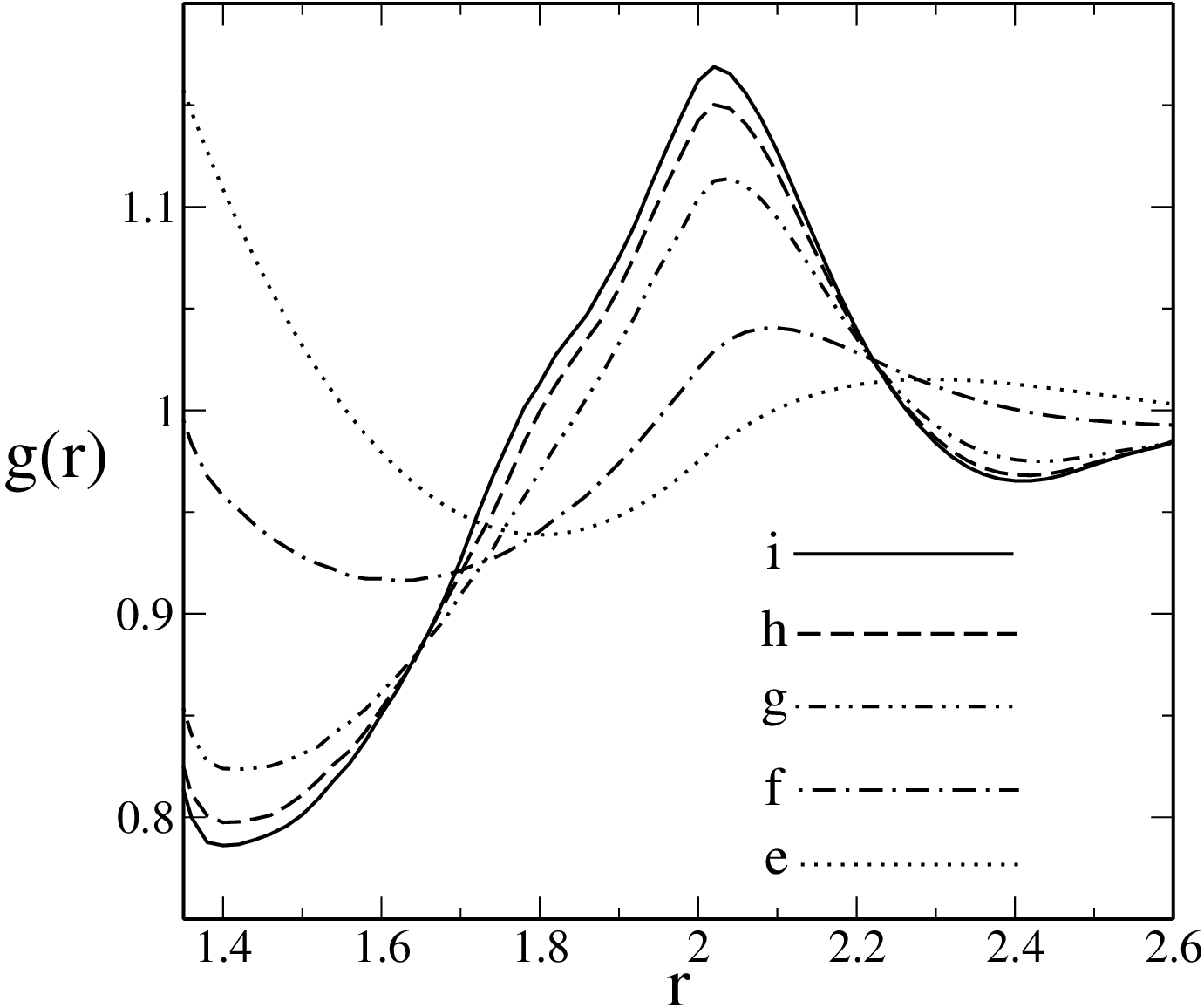}
\caption{}
\label{fig7}
\end{figure}

Figure 7. As for Fig.\ref{fig6} but for the 
statepoints labelled e-i in Fig.\ref{fig5}. In each case the packing fraction  
$\eta=0.20$. Statepoint (e) $\eta_p^r=0$ (dotted line), (f) $\eta_p^r=0.2$ 
(dot-dashed line), (g) $\eta_p^r=0.4$ (double dot-dashed line), 
(h) $\eta_p^r=0.45$ (broken line) and (i)  $\eta_p^r=0.47$ (full line).\newpage

\begin{thebibliography}{}
\bibitem{foot1}
We have performed numerical calculations for all of the approximate closures 
mentioned in the text and find the shoulder to be absent in all cases.
\bibitem{choudhury1}
The original Verlet-Modified closure \cite{verlet_80} was developed for hard spheres 
but proved less successful for systems with attractive interactions. 
The improved version (\ref{verlet_mod}) 
employed in this work was subequently proposed in \cite{choudhury} to correct 
these failings.
\end{thebibliography}
\end{document}